\documentclass[aps,prl,showpacs,twocolumn,superscriptaddress]{revtex4}

\usepackage{graphicx}
\usepackage{dcolumn}
\usepackage{bm}

\begin{document}
\newcommand{\maya}[1]{{\huge{$\bullet$}}{\em #1}}
\newcommand{\vadim}[1]{{\large{$\clubsuit$}}{\em #1}}
\newcommand{\alex}[1]{{\large{$\spadesuit$}}{\em #1}}


\title{Symbiotic Self-Organized Criticality and Intermittent Turbulence \\ 
in an MHD Current Sheet with a Threshold Instability}

\author{Alexander J. Klimas}%
\affiliation{NASA Goddard Space Flight Center, Greenbelt, MD 20771, USA}

\author{Vadim M. Uritsky}
\affiliation{Department of Physics and Astronomy, University of Calgary, Calgary, Alberta, Canada T2N 1N4}

\author{Maya Paczuski}%
\affiliation{Department of Physics and Astronomy, University of Calgary, Calgary, Alberta, Canada T2N 1N4}

\date{\today}

\begin{abstract}

We report numerical evidence of a self-organized criticality (SOC) and intermittent turbulence (IT) symbiosis in a resistive magnetohydrodynamic (MHD) current sheet model that includes a local hysteretic switch to capture plasma physical processes outside of MHD that are described in the model as current-dependent resistivity. Results from numerical simulations show scale-free avalanches of magnetic energy dissipation characteristic of SOC, as well as multi-scaling in the velocity field numerically indistinguishable from certain hierarchical turbulence theories. We suggest that SOC and IT may be complementary descriptions of dynamical states realized by driven current sheets -- which occur ubiquitously in astrophysical and space plasmas.

\end{abstract}

\maketitle

Most theoretical studies of self-organized criticality (SOC) focus on cellular models such as the paradigmatic BTW sandpile ~\cite{Bak87, Bak88}.  On the other hand, many examples of SOC-like phenomena occur in systems whose canonical descriptions invoke continuum equations such as the Navier-Stokes or magnetohydrodynamic (MHD) equations, whose solutions can exhibit some properties of intermittent turbulence (IT). It has been argued that complementarity between SOC and IT can be realized in avalanching systems~\cite{Bak05, Chen03, Chen04, Sreev04, Maya05, UP06, Chang99}.  The proposal that SOC and IT could be distinguished by analyzing waiting times between bursts~\cite{Boffetta} has turned out to be false: Once a finite observational threshold (unavoidable in any physical measurement) is introduced, even the ordinary BTW sandpile exhibits scale free waiting time statistics ~\cite{Maya05}. Simultaneous signatures of SOC and IT have been observed in physical systems such as, e.g., the solar corona through an analysis of an extended set of high resolution images provided by the SOHO spacecraft~\cite{UP06}.  Earth's magnetosphere produces power-law dissipative event statistics ~\cite{Urit02, Urit03, Urit06} as well as related intermittent plasma turbulence~\cite{Ang99, Lui06, Vor06}.  Avalanches with signatures of critical behavior have also been detected in numerical fluid models ~\cite{Dmit97, Klim04} but the potential complementarity between SOC and IT has not been rigorously proved.

Here, we analyze a current sheet model~\cite{Klim04,Klim05} that supports a magnetic field reversal in an MHD plasma. The model contains measurable plasma parameters coupled through the full system of resistive MHD equations, with boundary conditions that allow a persistent current sheet to form.  To describe local breakdown, where standard MHD equations fail and the true plasma equations must be invoked, we adapt the idea of E.~Lu \cite{Lu95}. A current dependent threshold instability, taking the form of an hysteretic switch controlling plasma resistivity~\cite{Klim04,Klim05}, is used.  With a uniform drive, the model repeatedly cycles through large scale loading and unloading phases.  We observe simultaneous signatures of SOC and IT in the dynamics of the model during the unloading phase. We limit our discussion to two of these: (1) We show that the model exhibits scale-free avalanching, consistent with SOC, of magnetic energy into a central reconnection zone where the field is dissipated, and (2) we provide evidence that this avalanching drives fluctuations in the velocity field that exhibit multi-scaling indistinguishable from certain hierarchical models of IT ~\cite{She94,Grauer94}.

Klimas et al. [2004, 2005] have shown that the presence of the hysteretic switch in the model leads to the propagation of micro-scale current waves that induce concurrent waves of resistivity, thus leading to local unfreezing of the magnetic field to form field energy avalanches with scale-free statistics. Here we provide evidence that localized, intermittent $\bm{J}\times\bm{B}$ acceleration associated with these current waves is directly responsible for the multi-scaling velocity field. In this current-sheet model, the SOC and IT dynamics are intimately associated in this way. We have found it impossible to have one without the other.

The 2d numerical model discussed here was originally motivated by observations of Earth's magnetotail, which is a driver for high-latitude geomagnetic activity~\cite{Baker99}.   The key plasma structure that controls geomagnetic substorms and is represented by our model is the large-scale cross-tail current sheet. The model consists of the full compressible, resistive MHD system, including a polytropic energy equation (see eq.(1) - (4) in~\cite{Klim04} for a complete  definition). In addition the local resistivity, $D$, obeys the following equations:
\begin{eqnarray}
Q= \left\{ 
\begin{array}{c}
D_{min}, |J| < \beta J_c; \ \ 
D_{max}, |J| > J_c \ \
\end{array}
\label{eq5}
\right\}
\\
\partial_t D=(Q(|J|)-D)/\tau \quad ,
\label{eq6}
\end{eqnarray}
which were adapted from Lu~\cite{Lu95}.  The function $Q$ takes the value $D_{max} \approx 1$ wherever the local current density $J$ exceeds a critical value $J_c$ and returns to $D_{min} \ll D_{max}$ when the current density falls below $\beta J_c$, with $\beta < 1$. Hence, the switch in $Q$ is hysteretic.  The transition from $D_{min}$ to $D_{max}$ represents the excitation and saturation of a kinetic current-driven instability over a time interval that is below the resolution of MHD and, hence, enters as an instantaneous transition. The transition from $D_{max}$ to $D_{min}$ represents the subsequent quenching of the instability.  The resistivity, $D$, is assumed to grow or decay with a single time-scale $\tau$ that is slow compared to the simulation time step. The values of the model parameters used here can be found in~\cite{Klim04}.

We have obtained numerical solutions of the current sheet model on a $400 \times 400$ grid with $D_{min} \ll 1$ such that wherever $D=D_{min}$ the evolution is indistinguishable from ideal   MHD over the observed time scales. A configuration of the model in a snapshot during an unloading phase is illustrated in Fig.~1. Plasma inflowing at the upper and lower boundaries carries magnetic flux with it, thus increasing the strength of the magnetic field reversal (and hence the electric currents in the current sheet). Eventually the current reaches $J=J_c$ somewhere -- not necessarily at $z=0$ -- which starts an avalanche of magnetic flux transport toward the central region where flux is annihilated and converted to kinetic and thermal energy. This conversion process also drives plasma out of the region through the open boundary at the right.  A small portion ($10^{-3}$ - $10^{-2}$) of the input magnetic energy is carried out through this boundary as well.

\begin{figure}[htbp]
\vskip -2.5 cm
\includegraphics[width=8.5cm]{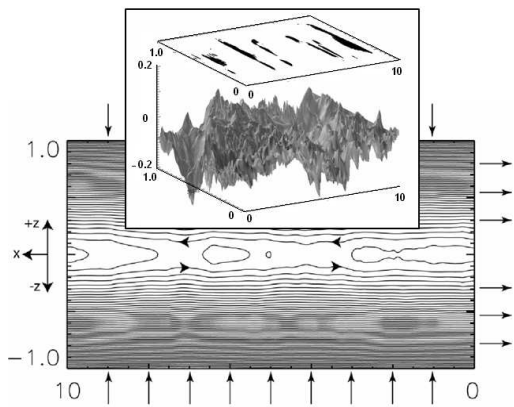}
\vskip -3 cm
\caption{\label{Fig_1} Numerical simulation of a current sheet and its
  qualitative behavior (see \cite{Klim04} for details). The model is
  driven by a steady, uniform plasma inflow at the top and bottom
  boundaries as shown by arrows.  The left boundary is closed,
  the right  is open.  Plasma energized through annihilation
  in the magnetic field reversal leaves the region at the right.
  Inset: (Top) A snapshot of regions where the diffusive Poynting flux
  exceeds a threshold, used to define avalanches. (Bottom)
  At the same time, the corresponding velocity field ($v_x$) in
  the system.  Note that while the magnetic field lines appear smooth
  at this scale, the plasma velocity field is highly intermittent.}
\vskip 0.0 cm
\end{figure}

Eventually, the simulated plasma reaches a statistically stationary state in which the rates of magnetic energy and plasma mass flowing into the region are balanced, over long time scales, by the field annihilation rate and the outflow at the open boundary.  After about $10^3$ Alfven traversal times, this state takes the form of {\it large scale global cycles} consisting of long laminar periods during which plasma is loaded into the system but no active grid sites (with $Q = D_{max}$) are generated, followed by highly erratic unloading periods during which the magnetic field undergoes local transitions between frozen and unfrozen states analogous to stick-slip behavior of SOC models~\cite{Bak89, Olami92, Maya96, Ged05}.  Fig.~2 illustrates the initiation of one of these unloading periods.
Neglecting a small convective contribution, the magnetic energy transported when a site becomes active is given by the local diffusive Poynting flux $\bm{S_d}(x,z)=(c/4 \pi)\eta \bm{J}\times\bm{B}$ ~\cite{Klim04}, in which $\eta = 2\pi D/c^2 $ is the anomalous resistivity, and $c$ is the speed of light. Fig.~2 shows the initial expansion of a Poynting flux avalanche in back of an outward propagating current wave as well as the associated transition from laminar to turbulent plasma flow at and in the interior of the expanding wave.

To observe avalanches, we used an automated technique for detecting and tracing regions having grid sites with $S_d$ above a certain threshold. In analogy with avalanches in 2d sandpiles, we treat the events as 2+1 dimensional spatiotemporal objects.  Avalanches were identified by applying a floating activity threshold $S_{th}(t)=\left\langle S_d \right\rangle + k \cdot \sigma $ adjusted to the average value $\left\langle S_d \right\rangle$ and the standard deviation $\sigma$ of the Poynting flux at every time step~\footnote{The floating   threshold allows improved statistics considering the   time variation  during unloading   cycles}.  The time evolution of avalanches was obtained by checking the intersections of spatial regions above $S_{th}(t)$ in consecutive pairs of $S_d(x,z)$ snapshots. Each avalanche was characterized by its lifetime $T$ and its total Poynting flux, $E$, obtained from the integration of $S_d$ over its spatiotemporal domain -- grid sites with $S_d(x,z,t) >S_{th}(t)$ taking part in the avalanche.  The linear dimensions $l_x$ and $l_z$ of avalanches were estimated by determining standard deviations of the $x$ and $z$ coordinates over all the grid sites involved in each avalanche (equally weighted). In addition the geometric mean $l_{xz} \equiv (l_x l_z)^{1/2}$, as well as the total area $s$, representing the total number of distinct pixels involved in the avalanche, were estimated.  The statistics reported here were obtained using $k=3.0$ and have also been reproduced in the range $k=1.5 - 4.0$.

\begin{figure}[htbp]
\includegraphics[width=\columnwidth] {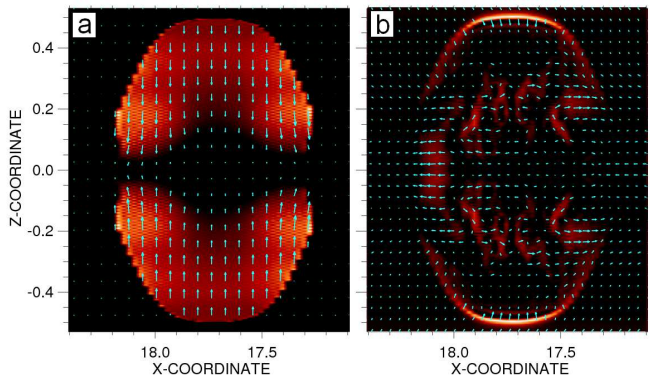}
\caption{\label{fig_2} The neighborhood of the initial site of instability shortly after the initiation of an unloading event. (a) Poynting flux due to slipping magnetic flux in the interior of the outward expanding current wave. (b) Transition from laminar to turbulent velocity field due to $\bm{J}\times\bm{B}$ acceleration at the current wave as it propagates through the magnetic field.}  
\end{figure}

The probability distribution for lengths, time, area, and energy of avalanches all obey scale free statistics.  The first group of critical exponents was estimated based on analyses of probability distributions $p(T,s_{max})$ and $p(E,s_{max})$ constructed from subsets with $s \leq s_{max}$, in which $s_{max}$ is defined to be the maximum area (number of pixels) of events included in the subset used to make the histogram. The normalized probability distributions were studied using the scaling {\it ansatz}
\begin{equation}
p(X,s_{max}) = X^{-\tau_X} f_X (X/X_c), \,\,\, X_c \sim s_{max}^{\lambda_X} 
\label{eq11}
\end{equation}
where $X \in \left\{ E,T\right\} $ and $f_X$ are  scaling functions that are approximately constant for $X < X_c$ and drop rapidly for $X>X_c$. Assuming Eq.~\ref{eq11}, we have plotted the distributions in the rescaled coordinates $\left(X/s_{max}^{\lambda_X},   p(X)X^{\tau_X}\right)$ and identified the combination of $\tau_E$, $\tau_T$, $\lambda_E$ and $\lambda_T$ exponents that provides the best data collapse (Fig.~3). The resulting values $\tau_E=1.48 \pm 0.02$ and $\tau_T=1.95 \pm 0.03$ coincide with those reported earlier in~\cite{Klim04}.  The exponents $\lambda_E=1.47 \pm 0.03$ and $\lambda_T=0.68 \pm 0.04$ are consistent with the regression analyses for the expected values of energy and lifetime for avalanches with a given size $s$  $\langle E \rangle_s$ and $\langle T \rangle_s$, within  statistical error.  These values also preserve the scaling relation $\lambda_T(\tau_T-1)=\lambda_E(\tau_E-1)$.

\begin{figure}[htbp]
\vskip -1.0 cm
\hskip -1.0 cm
\includegraphics*[width=9cm]{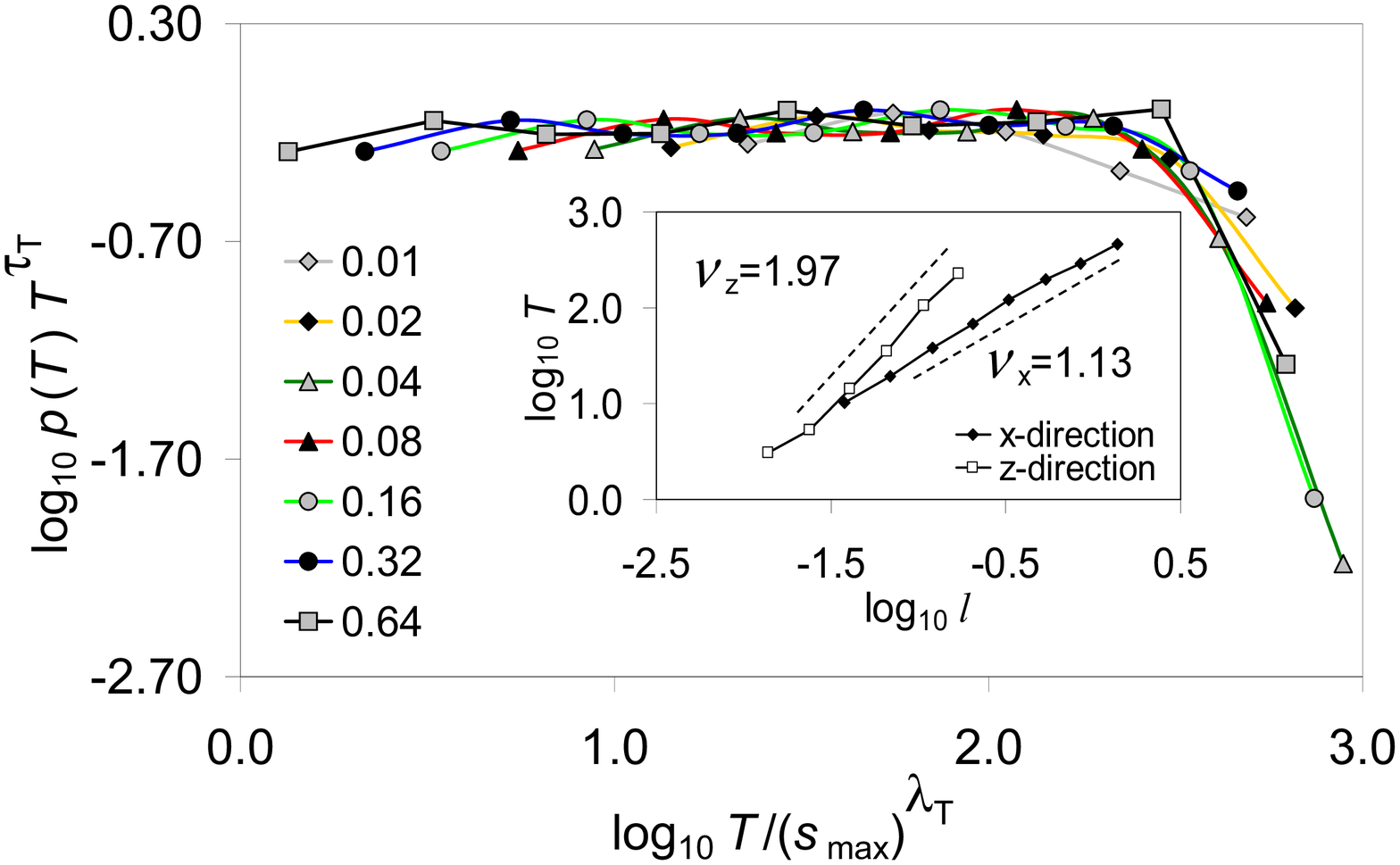}
\vskip -2.2 cm
\hskip -1.0 cm
\includegraphics*[width=9cm]{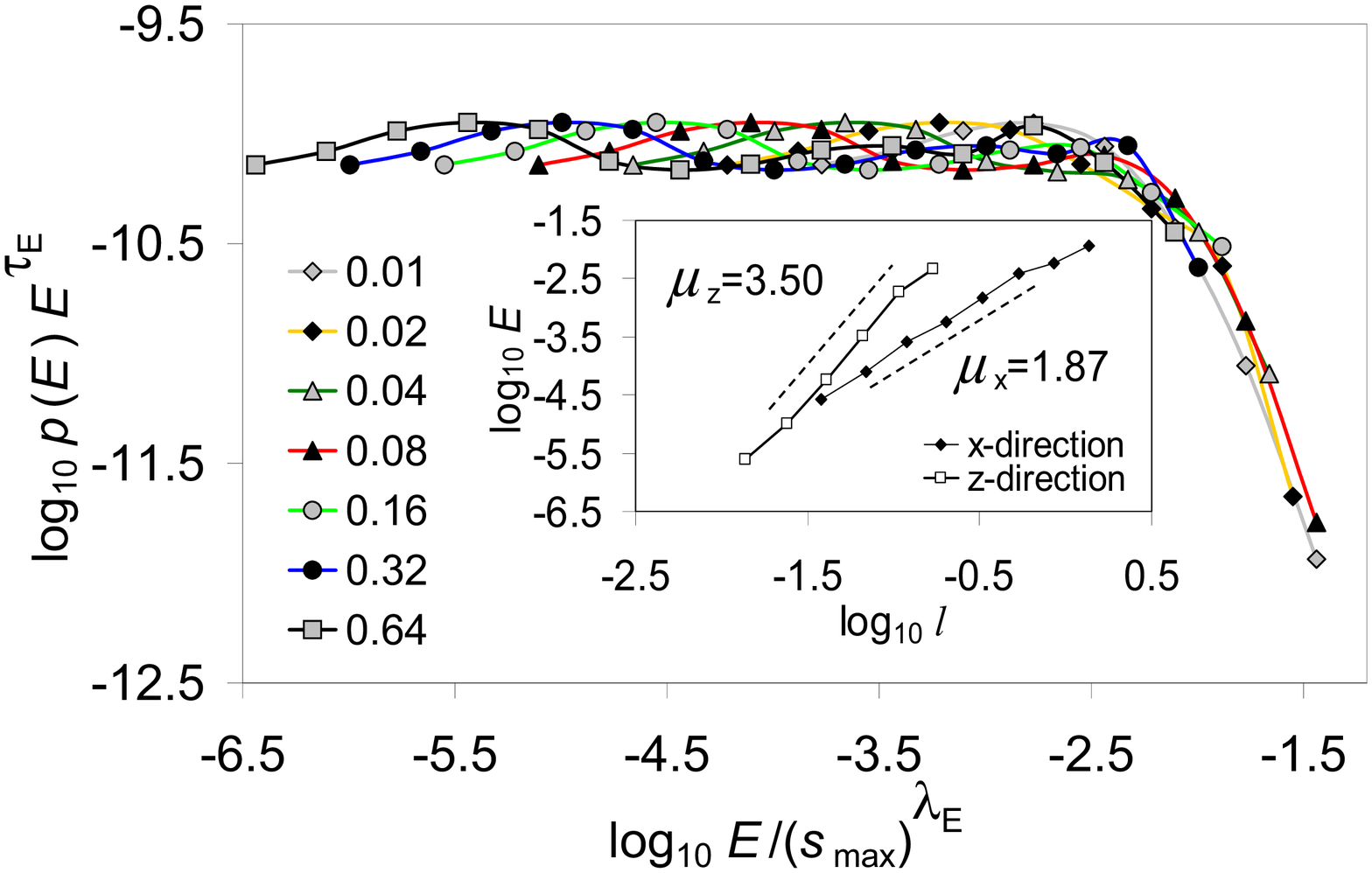}
\vskip -1.0 cm
\caption{\label{Fig_3} Data collapse using Eq. 3 for
  avalanches with different maximum area $s_{max}$ with $\tau_T=1.95$,
  $\lambda_T=0.68$, $\tau_E=1.48$, and $\lambda_E=1.47$. Insets:
  anisotropic scaling of $E$ and $T$ with the maximum avalanche extent $l$ in
  $x$ and $z$ directions.}
\vskip -0.3 cm
\end{figure}

The anisotropy of the model leads to different growth rates of avalanches in the $x$ and $z$ directions.  Hence  $\langle s\rangle_{l_x} \sim (l_x)^{d_x}$ and $\langle s\rangle_{l_z} \sim (l_z)^{d_z}$ with $d_x=1.40 \pm 0.03$ and $d_z=3.11 \pm 0.06$. The geometric mean $l_{xz}$ is related to $s$ through another scaling relation, $\langle s\rangle_{l_{xz}} \sim (l_{xz})^{d_{xz}}$, with $d_{xz}=1.97 \pm 0.05$ indicating that  avalanches are compact. The exponent values obtained are consistent with $d_x^{-1}+d_z^{-1} = 2 d_{xz}^{-1}$.  We also studied $E$ and $T$ as functions of $l_x$ and $l_z$ and found that $\langle E \rangle_{l_x} \sim l_x^{\mu_x}$, $\langle T\rangle_{l_x} \sim l_x^{\nu_x}$, $\langle E\rangle_{l_z} \sim l_z^{\mu_z}$ and $\langle T \rangle_{l_z}\sim l_z^{\nu_z}$ with scaling exponents $\mu_x = 1.87 \pm 0.04$, $\nu_x = 1.13 \pm 0.01$, $\mu_z = 3.50 \pm 0.04$ and $\nu_z = 1.97 \pm 0.06$ (see the insets in Fig.~3). The ratios $\mu_x/\mu_z$ and $\nu_x/\nu_z$ are close to $d_x/d_z$ as expected.  

All these results indicate that with respect to bursts of energy dissipation above background, the system operates at or near a SOC state. Based on the values of the distribution exponents $\tau_E$ and $\tau_T$ one can conjecture that the model operates near the mean-field limit. This is not typical for non-directed SOC sandpiles whose upper critical dimension $d_{u}$ is usually higher than 2. However, it is possible that the avalanching dynamics in our model can be mapped onto the universality class of non-Abelian directed sandpiles with irreversible topplings \cite{Hugh02} which exhibit the mean-field exponents starting from $d_u$=2. 

To analyze the velocity field, we have computed a set of equal time structure functions defined as  $S_q(l)= \left\langle |\delta v_l|^q \right\rangle$, where $\delta  v_l= (\bm{v}(\bm{r}+\bm{l}) - \bm{v}(\bm{r})) \cdot \bm{l}/l$ is the  increment of the velocity $\bm{v}$ in the direction $\bm{l}$  (parallel to $x$ or $z$ axes), $q$ is the order of the structure  function, $l \equiv |\bm{l}|$ is the spatial displacement, and  averaging indicated by $\langle \cdots \rangle$ is performed over all  positions $\bm{r}$ and times during an unloading phase.  For many turbulent phenomena, $S_q(l) \sim  l^{\zeta(q)}$ with $\zeta(q)$ defined by the turbulent regime under  study. To extend the scaling range and improve the accuracy of this analysis, we have  applied the method of extended self similarity (ESS)~\cite{Benz93} by  plotting $S_q(l)$ versus $S_3(l)$.

The resulting structure functions (Fig.~4) exhibit ESS over the  entire range of scales available. The error bars shown are for  $\zeta(q)/ \zeta(3)$  in  the $x$ direction; the errors in the $z$ direction are about three times  smaller.  The values obtained in both directions are the same up  to these errors. The dependence of $\zeta(q)/ \zeta(3)$ on the order  $q$ shows a systematic departure from the Kolmogorov law $\zeta=q/3$.  In principle, it can be fitted by the hierarchical model  $\zeta(q)=(1-\gamma)q/g + C(1-[1-\gamma/C]^{q/g} ]$, in which $C$ is  the codimension of the most singular dissipative structures, $g$ and  $\gamma$ are defined by $\delta z \sim \ell ^{1/g}$ and $t_e \sim  \ell ^\gamma$, with $t_e$ being the energy transfer time at the smallest  inertial scales $\ell$~\cite{Polit95}. By choosing either $g=4$,  $\gamma=1/2$, $C=1$ (Iroshnikov-Kraichnan theory (IK)~\cite{Grauer94}) or $g=3$,  $\gamma=2/3$, $C=2$ (She and Leveque (SL) theory ~\cite{She94}) one can obtain rather accurate fits to the data. However, the physical conditions for the turbulence in our model are strikingly different from those in any of these hierarchical models. The current waves at the leading fronts of energy avalanches accelerate the fluid through the $\bm{J}\times\bm{B}$ force. These current structures play the role of energy sources rather than energy sinks (as would be the case in classical turbulent models). As we have shown, the avalanches are scale-free, and thus this driving mechanism appears at all scales as opposed to the standard picture of the direct turbulent cascade. At this point there is no closed theory that could describe the relationship between the magnetic energy avalanches and the intermittency in the velocity field (e.g. in the form of an ``exact law'' \cite{Conn07}), although it is evident that the resulting behavior is different from the MHD turbulence ~\footnote{This is also evident from the 3d version of our model which will be described in a separate paper.}.

\begin{figure}[htbp]
\vskip -0.5 cm
\hskip -0.45 cm
\includegraphics*[width=5.1cm]{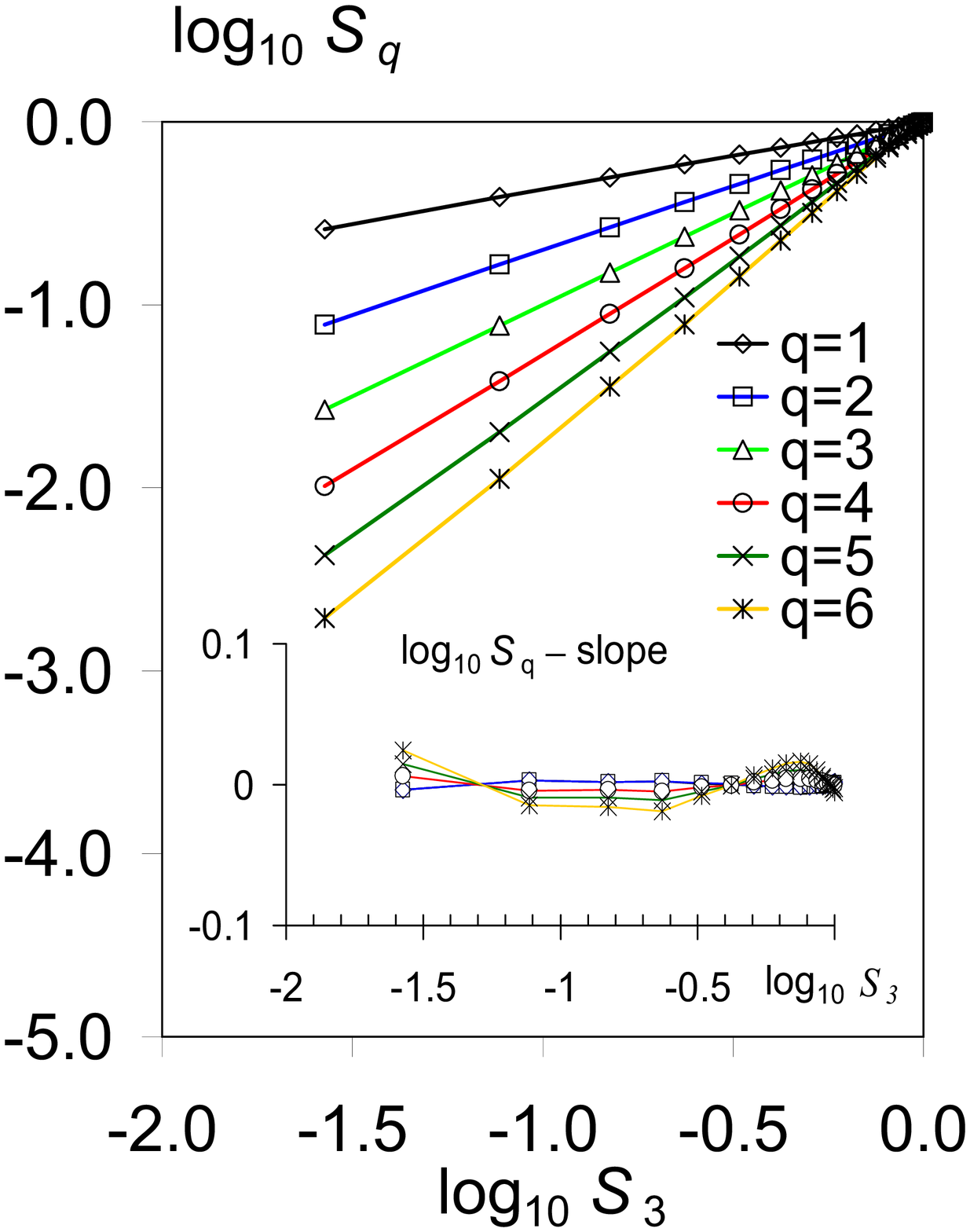}
\hskip -1.2 cm
\includegraphics*[width=5.1cm]{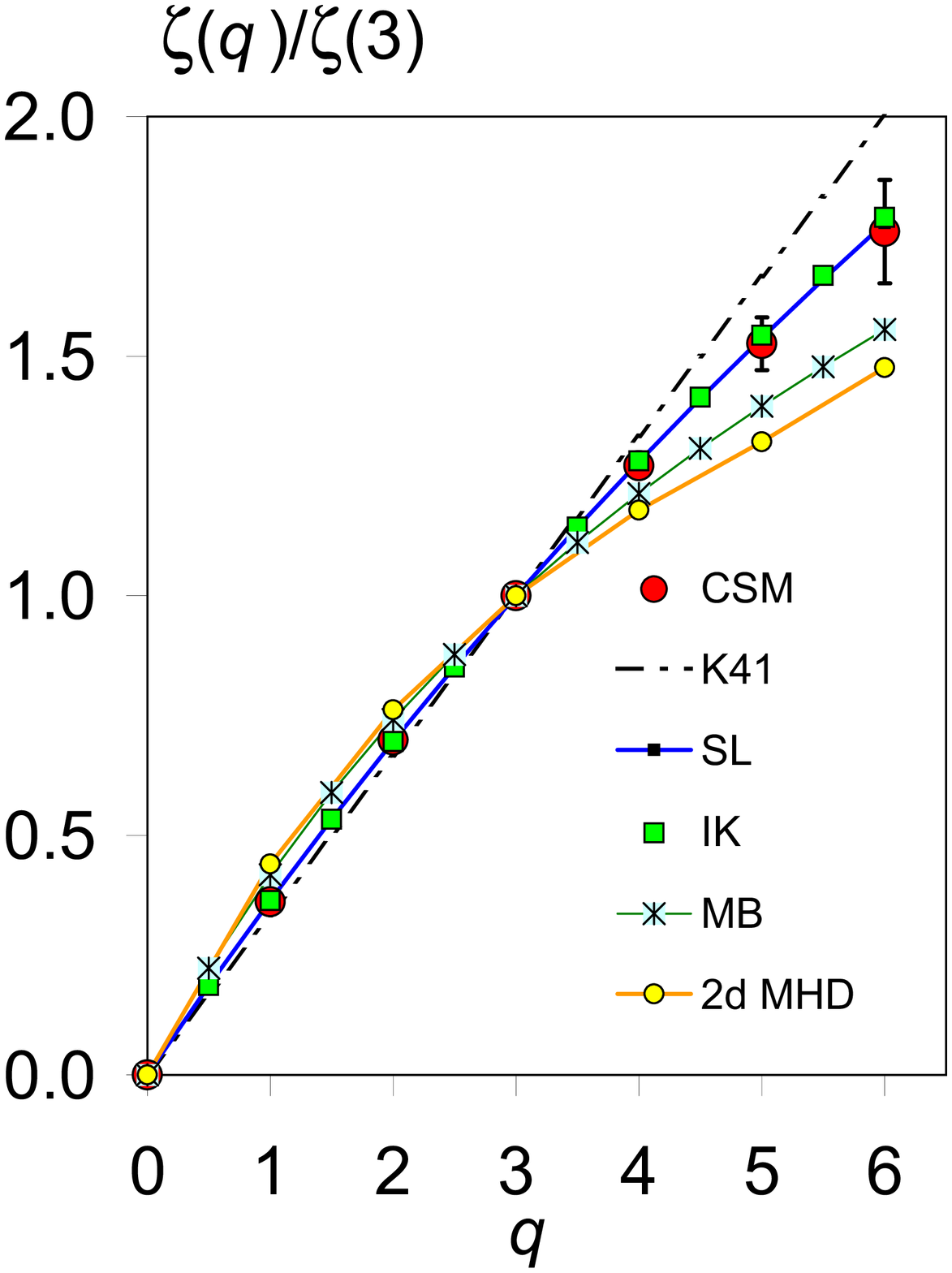}
\vskip -0.5 cm
\caption{\label{Fig_4} Left: ESS plots of velocity structure functions
with $l$ parallel to $z$ axis. The inset shows the same functions
with subtracted average slopes.
Right: Dependence of $\zeta(q)/ \zeta(3)$ on the order $q$ for our
current sheet model (CSM, with error bars), hierarchical models of IT mentioned in the text, M\"{u}ller and Biskamp  (MB)~\cite{Mull00} model ($g=3$, $\gamma=2/3$, $C=1$), Kolmogorov model (K41) \cite{Kolm41}, as well as the exponents from 2d ideal MHD simulations \cite{Polit98}. }
\vskip -0.0 cm
\end{figure}

The interplay between SOC- and turbulence-based routes to multiscale complexity in natural systems remains a subject of extreme interest in the modern physical literature. There is a growing body of evidence that statistical signatures of these routes coexist in real-world processes, including those in the solar corona and Earth's magnetosphere. At the same time, there is no clear understanding of how such coexistence arises from first principles of fluid dynamics. Our study partially fills this gap by providing an example of a realistic (and very common, see e.g.~\cite{Priest00, Asch05,Cowley97}) plasma configuration in which avalanches of electromagnetic energy stir an otherwise laminar velocity field to generate a self-consistent pattern of multiscale spatiotemporal fluctuations obeying both SOC and turbulent scaling laws. There is no doubt that the resulting behavior is a more general form of dynamical complexity than either of the two scenarios separately. We are looking forward to seeing future theoretical studies of this phenomenon. 


\vskip 0.1 cm

\begin{acknowledgments}
The work of A.~J.~K. was supported by the NASA Geospace Sciences program.
\end{acknowledgments}

\bibliography{AlexModel}

\end{document}